\PassOptionsToPackage{unicode}{hyperref}
\PassOptionsToPackage{hyphens}{url}
\PassOptionsToPackage{dvipsnames,svgnames,x11names}{xcolor}
\documentclass[
]{article}
\usepackage{amsmath,amssymb}
\usepackage{lmodern}
\usepackage{iftex}
\ifPDFTeX
  \usepackage[T1]{fontenc}
  \usepackage[utf8]{inputenc}
  \usepackage{textcomp} 
\else 
  \usepackage{unicode-math}
  \defaultfontfeatures{Scale=MatchLowercase}
  \defaultfontfeatures[\rmfamily]{Ligatures=TeX,Scale=1}
\fi
\IfFileExists{upquote.sty}{\usepackage{upquote}}{}
\IfFileExists{microtype.sty}{
  \usepackage[]{microtype}
  \UseMicrotypeSet[protrusion]{basicmath} 
}{}
\makeatletter
\@ifundefined{KOMAClassName}{
  \IfFileExists{parskip.sty}{%
    \usepackage{parskip}
  }{
    \setlength{\parindent}{0pt}
    \setlength{\parskip}{6pt plus 2pt minus 1pt}}
}{
  \KOMAoptions{parskip=half}}
\makeatother
\usepackage{xcolor}
\usepackage{graphicx}
\makeatletter
\def\maxwidth{\ifdim\Gin@nat@width>\linewidth\linewidth\else\Gin@nat@width\fi}
\def\maxheight{\ifdim\Gin@nat@height>\textheight\textheight\else\Gin@nat@height\fi}
\makeatother
\setkeys{Gin}{width=\maxwidth,height=\maxheight,keepaspectratio}
\makeatletter
\def\fps@figure{htbp}
\makeatother
\providecommand{\tightlist}{%
  \setlength{\itemsep}{0pt}\setlength{\parskip}{0pt}}
\NewDocumentCommand\citeproctext{}{}
\NewDocumentCommand\citeproc{mm}{%
  \begingroup\def\citeproctext{#2}\cite{#1}\endgroup}
\makeatletter
 \let\@cite@ofmt\@firstofone
 \def\@biblabel#1{}
 \def\@cite#1#2{{#1\if@tempswa , #2\fi}}
\makeatother
\newlength{\cslhangindent}
\newlength{\csllabelwidth}
\newenvironment{CSLReferences}[2] 
 {\begin{list}{}{%
  \setlength{\itemindent}{0pt}
  \setlength{\leftmargin}{0pt}
  \setlength{\parsep}{0pt}
  \ifodd #1
   \setlength{\leftmargin}{\cslhangindent}
   \setlength{\itemindent}{-1\cslhangindent}
  \fi
  \setlength{\itemsep}{#2\baselineskip}}}
 {\end{list}}
\usepackage{calc}

\ifLuaTeX
\usepackage[bidi=basic]{babel}
\else
\usepackage[bidi=default]{babel}
\fi
\babelprovide[main,import]{american}

\def\languageshorthands#1{}
\ifLuaTeX
  \usepackage{selnolig}  
\fi
\IfFileExists{bookmark.sty}{\usepackage{bookmark}}{\usepackage{hyperref}}
\IfFileExists{xurl.sty}{\usepackage{xurl}}{} 
\hypersetup{
  pdftitle={PiNNAcLe: Adaptive Learn-On-The-Fly Algorithm for
Machine-Learning Potential},
  pdfauthor={Yunqi Shao, Chao Zhang},
  pdflang={en-US},
  colorlinks=true,
  linkcolor={Maroon},
  filecolor={Maroon},
  citecolor={Blue},
  urlcolor={Blue},
  pdfcreator={LaTeX via pandoc}}

\title{PiNNAcLe: Adaptive Learn-On-The-Fly Algorithm for
Machine-Learning Potential}

\usepackage[affil-it]{authblk}
\usepackage{orcidlink}
\author[1,2%
  \ensuremath\mathparagraph]{Yunqi Shao%
    \,\orcidlink{0000-0002-5769-5558}\,%
    }
\author[1%
  ]{Chao Zhang%
    \,\orcidlink{0000-0002-7167-0840}\,%
    }

\affil[1]{Department of Chemistry-Ångström Laboratory, Uppsala
University, Lägerhyddsvägen 1, BOX 538, 75121 Uppsala, Sweden}
\affil[2]{Chalmers eCommons, Chalmers University of Technology, 41296
Gothenburg , Sweden}
\affil[$\mathparagraph$]{Corresponding author}
\date{2 September 2024}

\begin{document}
\maketitle

\section{Summary}\label{summary}

PiNNAcLe is an implementation of our adaptive learn-on-the-fly algorithm
for running machine-learning potential (MLP)-based molecular dynamics
(MD) simulations -- an emerging approach to simulate the large-scale and
long-time dynamics of systems where empirical forms of the PES are
difficult to obtain.

The algorithm aims to solve the challenge of parameterizing MLPs for
large-time-scale MD simulations, by validating simulation results at
adaptive time intervals. This approach eliminates the need of
uncertainty quantification methods for labelling new data, and thus
avoids the additional computational cost and arbitrariness thereof.

The algorithm is implemented in the NextFlow workflow language
(\citeproc{ref-2017_DiTommasoChatzouEtAl}{Di Tommaso et al., 2017}).
Components such as MD simulation and MLP engines are designed in a
modular fashion, and the workflows are agnostic to the implementation of
such modules. This makes it easy to apply the same algorithm to
different references, as well as scaling the workflow to a variety of
computational resources.

The code is published under BSD 3-Clause License, the source code and
documentation are hosted on Github. It currently supports MLP generation
with the atomistic machine learning package PiNN
(\citeproc{ref-2020_ShaoHellstroemEtAl}{Shao et al., 2020}), electronic
structure calculations with CP2K
(\citeproc{ref-2020_KuhneIannuzziEtAl}{Kühne et al., 2020}) and DFTB+
(\citeproc{ref-2020_HourahineAradiEtAl}{Hourahine et al., 2020}), and MD
simulation with ASE (\citeproc{ref-2017_LarsenMortensenEtAl}{Larsen et
al., 2017}).

\section{Statement of need}\label{statement-of-need}

Recent development of MLPs see great progress in the understanding
complete descriptions of atomic configurations,
(\citeproc{ref-2021_MusilGrisafiEtAl}{Musil et al., 2021}) underpinning
the feasibility of parameterizing MLPs at high accuracy and low
computational cost. That said, generation of MLPs is still challenging
since reference data are scarce and their distribution is unknown to the
primary application of MLPs -- to sample statistic distributions where
labelling is expensive.

A full understanding of MLPs thus calls for understanding the
parameterization procedure, the following aspects:

\begin{enumerate}
\def\labelenumi{\arabic{enumi}.}
\tightlist
\item
  The coupling between label generation and sampling with MLPs, i.e.~how
  limited initial data limit the extrapolation ability of MLPs, which in
  turn affects the data generation;
\item
  The most representative distribution of data given a specific system
  or dataset, i.e.~how procedures like query-by-committee affects the
  performance of MLPs; (\citeproc{ref-2018_SmithNebgenEtAl}{Smith et
  al., 2018}; \citeproc{ref-2021_Zaverkin}{Zaverkin \& Kästner, 2021})
\item
  How do hyperparameters of the MLPs, such as the architecture or
  training algorithm, influence the performance of MLPs
  (\citeproc{ref-2021_ShaoDietrichEtAl}{Shao et al., 2021})
\end{enumerate}

While 2 and 3 are method-specific, 1 is a general problem for the
development of all MLPs and it precedes 2 and 3 in most cases. Classical
learn-on-the-fly (LOTF) and active learning (AL) are two available
paradigms to tackle the data-generation challenge mentioned above.

In classical LOTF, labelling is carried out at a fixed time interval and
MLPs were used to extrapolate for 5 to 30 molecular dynamics (MD) steps
in between (\citeproc{ref-2004_CsanyiAlbaretEtAl}{Csányi et al., 2004};
\citeproc{ref-2015_LiKermodeDeVita}{Li et al., 2015}). This has the
clear limitation that reference calculations must be continuously
carried out during the simulation.

In AL, the performance of MLPs during the extrapolative sampling is
gauged with an uncertainty estimation, through e.g.~Bayesian inference
(\citeproc{ref-2019_JinnouchiKarsaiEtAl}{Jinnouchi et al., 2019};
\citeproc{ref-2020_VandermauseTorrisiEtAl}{Vandermause et al., 2020}),
D-optimality (\citeproc{ref-2017_PodrybinkinEvgenyEtAl}{Podryabinkin \&
Shapeev, 2017}), or the query-by-committee strategy
(\citeproc{ref-2017_GasteggerBehlerMarquetand}{Gastegger et al., 2017};
\citeproc{ref-2020_SchranBrezinaEtAl}{Schran et al., 2020};
\citeproc{ref-2020_ZhangWangEtAl}{Zhang et al., 2020}). One can in
principle minimize the frequency of (and eventually skip) labelling by
focusing on the extrapolated data points estimated with large
uncertainty.

However, the assumed correlation between the test error and the
uncertainty estimation is only verified for very short MD runs
(\citeproc{ref-2015_Behler}{Behler, 2015};
\citeproc{ref-2019_JinnouchiKarsaiEtAl}{Jinnouchi et al., 2019}) and not
guaranteed for either the long timescale dynamics or the
out-of-distribution samples (\citeproc{ref-2018_UtevaGrahamEtAl}{Uteva
et al., 2018};
\citeproc{ref-2022_VazquezSalazarLuisEtAl}{Vazquez-Salazar et al.,
2022}; \citeproc{ref-2021_Zaverkin}{Zaverkin \& Kästner, 2021}). This
calls for an alternative parameterization procedure by introducing a
direct feedback mechanism to the classical LOTF without involving the
uncertainty estimation, which we name as \emph{the adaptive LOTF}.

\section{Adaptive LOFT algorithm}\label{adaptive-loft-algorithm}

\begin{figure}
\centering
\includegraphics[width=4cm,height=\textheight]{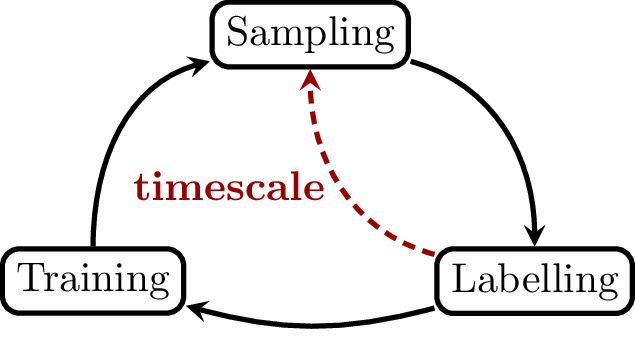}
\caption{Schematics of the adaptive LOTF scheme. Compared to the typical
feedback loop in either classical LOTF or AL, the labelling process in
adaptive LOTF directly affects the sampling timescale.\label{fig1}}
\end{figure}

Iterative workflow are common in applications of MLPs. What makes the
adaptive LOTF protocol distinct is how the timescale of the sampling
processes are adjusted for generation according to the convergence of
test set error in the previous generation, as highlighted in
\autoref{fig1}.

\begin{figure}
\centering
\includegraphics[width=8cm,height=\textheight]{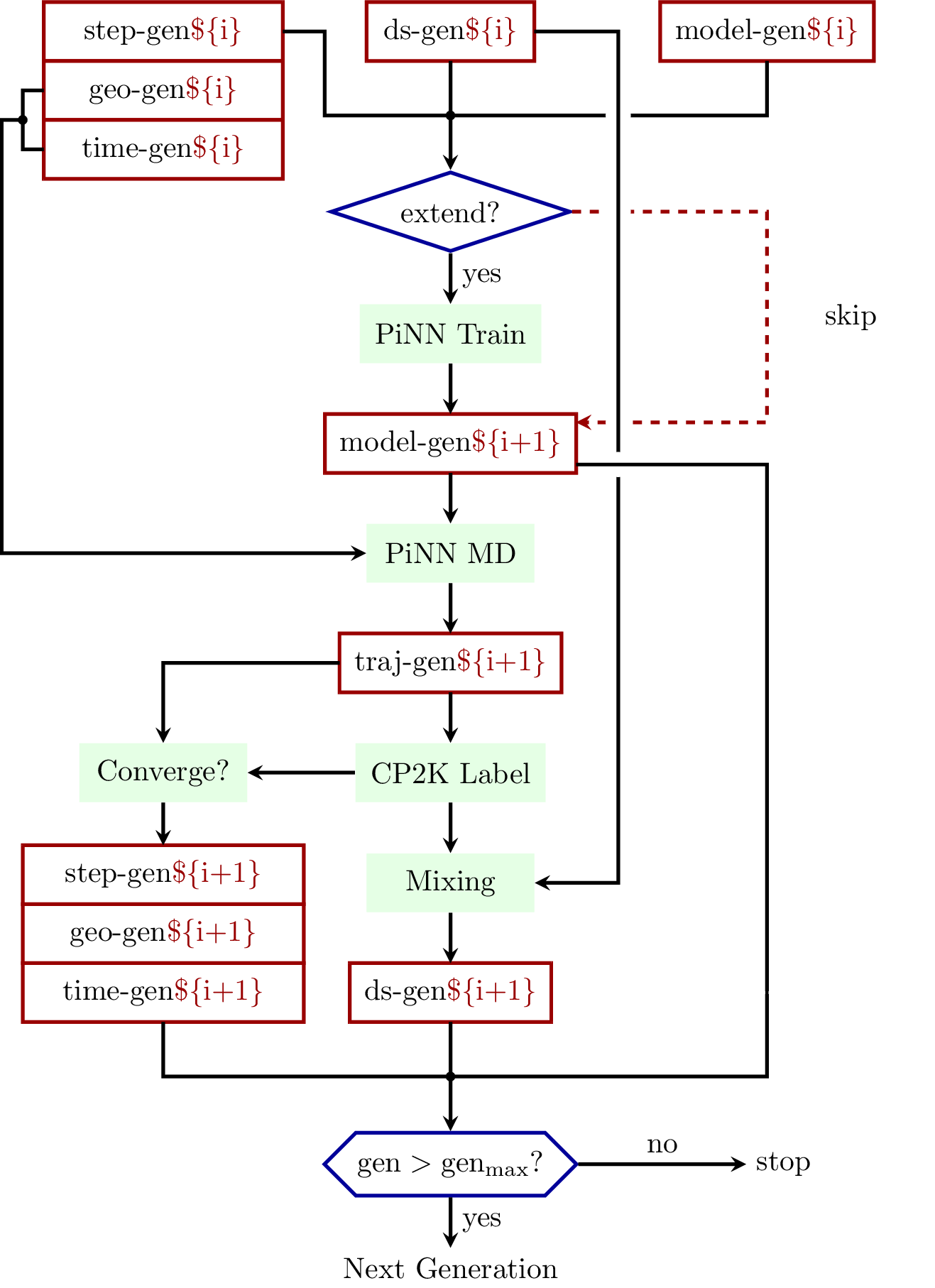}
\caption{Flowchart of the adaptive LOTF algorithm, Green boxes denote
processes, red boxes denote data which are updated at each iteration,
and blue boxes denote loops and decisions.\label{fig2}}
\end{figure}

The adaptive LOTF algorithm can be illustrated in terms of different
data and the transformation thereof, as shown in \autoref{fig2}.

The iterative sampling and training of NNP consists of a training
process (``PiNN Train''), a sampling process (``PiNN MD''), and a
labelling process (``CP2K Label''). Two auxiliary processes
(``Converge'' and ``Mixing'') checks the error of the trajectories and
generates a new training set. We denote such an iteration as a
``generation''. In each generation, the model, the dataset, and the
training step are updated for ``PiNN Train''; the initial geometry for
the extrapolative sampling and the corresponding sampling time are
updated for ``PiNN MD''; the snapshots taken from the extrapolative
sampling are updated for ``CP2K Label''.

Specifically, the convergence of each sampled trajectory is validated
against references. If all sampled trajectories are deemed as converged,
the ``PiNN MD'' sampling time (``time'') is multiplied by a factor for
the next generation, starting from a given small timescale to generation
to a maximal timescale where the sampling is deemed as sufficient.
Otherwise, ``PiNN Train'' in the next generation to incorporate new
reference data.

\section{Acknowledgements}\label{acknowledgements}

This work has been supported by the European Research Council (ERC)
under the European Union's Horizon 2020 research and innovation
programme (grant agreement No.~949012). The authors are thankful for the
funding from the Swedish National Strategic e-Science program eSSENCE,
STandUP for Energy and BASE (Batteries Sweden). The simulations were
performed on the resources provided by the Swedish National
Infrastructure for Computing (SNIC) at C3SE and PDC.

We thank A. Laio for reading the manuscript and for providing helpful
feedback. Y.S. also thanks A. Laio for hosting a research visit at the
International School for Advanced Studies (SISSA) and D. Doimo for many
useful discussions.

\section*{References}\label{references}
\addcontentsline{toc}{section}{References}

\phantomsection\label{refs}

\end{document}